\documentclass[prc,unsortedaddress,superscriptaddress,twocolumn,amssymb,showpacs,floatfix,nofootinbib]{revtex4-1}
\usepackage{graphics,epsfig}
\usepackage{amsmath}
\usepackage{dcolumn}   
\begin{document}
\date{\today}
\title{Asymptotic normalization coefficients of resonant and bound states from the phase shifts for $\alpha\alpha$ and $\alpha^{12}\rm C$ scattering.}
\author{Yu.V. Orlov}\email{orlov@srd.sinp.msu.ru}
\affiliation{Skobeltsyn Nuclear Physics Institute, Lomonosov Moscow State
University, Russia}
\author{B.F. Irgaziev}\email{irgaziev@yahoo.com}
\affiliation{Institute of Applied Physics, National University of Uzbekistan, Tashkent,Uzbekistan}
\affiliation{GIK Institute of Engineering Sciences and
Technology, Topi, Pakistan}
\author{L.I. Nikitina}
\affiliation{Skobeltsyn Nuclear Physics Institute, Lomonosov Moscow State
University, Russia}
\begin{abstract}
Recently we have published a paper [Irgaziev, Phys. Rev. C 91, 024002 (2015)] where the $S$-matrix pole method (SMP) which is only valid for resonances has been developed to derive an explicit expression for the asymptotic normalization coefficient (ANC), and is  applied to the  low-energy resonant states of nucleon$+\alpha$ and $\alpha+^{12}\rm{C}$ systems. The SMP results are compared with the effective-range expansion method (EFE) results. In the present paper the SMP and EFE plus the Pad\'{e}-approximation are applied to study the excited 2$^+$ resonant states of $^{8}\rm{Be}$. A contradiction is found between descriptions of the experimental phase shift data for $\alpha\alpha$ scattering and of the $^{8}\rm{Be}$ resonant energy for 2$^+$ state. Using the EFE method, we also calculate the ANC for the $^{8}\rm{Be}$ ground 0$^+$ state with a very small width. This ANC agrees well with the value calculated using the known analytical expression for narrow resonances.  In addition, for the $\alpha+^{12}\rm{C}$ states 1$^-$ and 3$^-$ the SMP results are compared with the Pad\'{e} approximation results. We find that the Pad\'{e}-approximation improves a resonance width description compared with the EFE results. The EFE method is also used to calculate the ANCs for the bound $^{16}\rm{O}$ ground 0$^+$ state and for the excited 1$^-$ and 2$^+$ levels which are situated near the threshold of $\alpha+^{12}\rm{C}$ channel.
\end{abstract}
\maketitle

\section{Introduction}\label{intr}
It is well-known that nuclear reactions at very low energies are key nuclear processes in stellar nucleosynthesis \cite{imbriani, rausher, straniero, hill}. These reactions can occur as direct processes or through the formation of resonance. A direct reaction near the threshold is suppressed due to the Coulomb barrier, and the reaction through resonance therefore becomes critical when it is near the threshold. Examples of certain resonances of interest to nuclear astrophysics are presented below. In Ref. \cite{irgaz} we developed the $S$-matrix pole method (SMP) to derive an explicit expression for the asymptotic normalization coefficient (ANC) of a resonant Gamow radial wave function in the presence of the Coulomb interaction. We used an analytical $S$-matrix approximation in the form of a series of powers of the relative momentum $k$ for the nonresonant part of the phase shift for the arbitrary orbital momentum $l$ of colliding particles. Earlier in Ref. \cite{akram09}, this method was applied to calculate the resonance pole energy $E_r$ and width $\Gamma$. In the earlier paper \cite{irgaz} we successfully applied the SMP to the resonances of the nucleon$+\alpha$ and $\alpha+^{12}\rm{C}$ systems. We obtained respective ANC values for different states of the nuclei $^5\rm\rm{He}, ^5\rm{Li}$, and $^{16}\rm{O}$ . We found that the $\Gamma$ value calculated by the SMP agrees better with the other values from the literature, while the effective-range expansion method (EFE) overestimates $\Gamma$.

In the present paper we apply low-energy approaches (SMP, EFE and Pad\'{e}-approximant) to the  another nucleus $^8\rm{Be}$ which is unstable even in the ground $0^+$ state.

Hoyle predicted the existence of the resonance state of the $^{12}\rm{C}$ nucleus with an excitation energy of 7.68 MeV \cite{Hoyle} even before the actual observation of 7.65 MeV in experiments.  Salpeter accepted Hoyle�'s idea and theoretically considered the $^{12}\rm{C}$ creation mechanism as the result of a three $\alpha$ particles fusion with the intermediate creation of the narrow resonance $^{8}\rm{Be}$ in the ground state. Fowler and his group carried out corresponding experiments which confirmed Hoyle's  prediction. At the end of the life of red giant stars compressed by the gravitation, the temperature increases up to values $T > 10^8$ K. At such temperatures carbon creation occurs due to the two consecutive processes: $\alpha+\alpha\rightarrow\, ^{8}\rm{Be}$ ( 0$^{+}$, ground state) and $\alpha +\, ^{8}\rm{Be}\rightarrow \,^{12}\rm{C}$* ($0^+$, 7.65 MeV). The small difference ($\cong$ 0.28 MeV) between the $^{12}\rm{C}$ energy level and those of the system $\alpha +  \,^{8}\rm{Be}$ is especially important.

The EFE is also used in the present paper to calculate ANCs for the bound states of the system $\alpha+\,^{12}\rm{C}$. Besides, we find that it is possible to improve an agreement for the resonance width $\Gamma$ when  a Pad\'e-approximant is used instead of the EFE. In our previous paper \cite{irgaz} we study $^{16}\rm{O}$ resonances in $\alpha^{12}\rm{C}$ scattering. However, the properties of the $^{16}\rm{O}$ bound states are also studied here because they are quite important for astrophysics.

The $^{12}\rm{C}(\alpha,\gamma)^{16}\rm{O}$ reaction is considered one of the key nuclear processes at the stage of helium combustion in stars (formation of the red giant). This reaction determines the relative content of $^{12}\rm{C}/^{16}\rm{O}$ in the process of stellar helium combustion. It directly affects the sequence and peculiarities of further combustion stages in massive stars, such as carbon combustion. One needs to know the ANC for the decay $^{16}\rm{O}\rightarrow \alpha+\,^{12}$C for different final states to find the rate of this radiative capture reaction.
The main goals of the present paper are the applications  of low-energy approaches to the other nucleus $^8\rm{Be}$ and to the $^{16}\rm{O}$ states which are of interest to astrophysics but are not investigated in Ref. \cite{irgaz}. The article is organized as follows. In Sec. \ref{main} we present main formulas of the $S$-matrix pole method.  In Sec. \ref{alpha-alpha} we study the properties of the ground $^8\rm{Be}$ $s$ wave state resonance in the $\alpha\alpha$ scattering. Short history of this level complex energy studying in the literature is given as well as our results for the energy, the nuclear vertex ($^8\rm{Be}\rightarrow\alpha+\alpha$) constant (NVC) and the ANC calculated using the EFE method. We have to use here the EFE instead of the SMP because this state situated just near the zero energy where the phase shift has essential singularity (see Eq. (20) in Ref. \cite{irgaz}). In Sec. \ref{alpha-alpha d state} we use all the three methods (SMP, EFE, and Pad\'{e}) to calculate the excited $d$ wave state resonance in the $\alpha\alpha$ scattering. We compare the SMP approximation results for the same excited $2^+$ resonant state of $^8\rm{Be}$  for the two variants when fitting the experimental phase shift data: taking the experimental resonance energy $E_2$ as fixed and considering $E_2$  as an additional fitting quantity. The latter leads to a very good agreement with the experimental phase shift data. It is shown that there is disagreement between the experimental energy dependence of the phase shift on the one hand and the resonance complex energy on the other. In Sec. \ref{alpha-12C} we consider the $^{16}\rm{O}$ bound  states: the ground $0^+$ state as well as the excited bound states $1^-$ and $2^+$. These states are very important for astrophysics. In Sec. \ref{alpha-12Cpade} we study the $^{16}\rm{O}$ resonant  states $1^-$ and $3^-$. with bigger widths, which was recently considered in Ref. \cite{irgaz} where the very good description of the experimental phase shift data was achieved. We show that the Pad\'{e} approximant describes width better than EFE. In Sec. \ref{conclusion} (conclusion) the results of the present paper are discussed.  For the bound $^{16}\rm{O}$ excited state $2^+$ our calculated ANC is compared with the one published in the literature.

\section{THE MAIN FORMULAS OF THE $S$-MATRIX POLE METHOD (SMP)}\label{main}

All the SMP formulas needed to calculate the resonance energy, nuclear vertex constant (NVC) and asymptotic normalization coefficient (ANC) are given  in Ref. \cite{irgaz}. Some of them are given below. \footnote{Here and below we use the unit system $\hbar=c=1$.} In the single-channel elastic
scattering case the partial $S$-matrix element (without the pure
Coulomb part $e^{i2\sigma_l}=\Gamma(l+1+i\eta)/\Gamma(l+1-i\eta)$ where $\Gamma(x)$ is the gamma-function, $\eta=z_1z_2\mu \alpha/k$ is the Sommerfeld parameter, $\alpha$ is the fine-structure
constant, and $\mu$ is the reduced mass of the colliding nuclei with the charge numbers $z_1$ and $z_2$)
is denoted as
\begin{equation}\label{S-l}
S_l(k)=e^{i2\delta_l}.
\end{equation}
Near an isolated resonance it can be approximated as \cite{migdal}
\begin{equation}\label{S-mat}
S_l(k)=e^{2i\nu_l(k)}\frac{(k+k_r)(k-k_r^\star)}{(k-k_r)(k+k_r^\star)},
\end{equation}
where $k_r=k_0-ik_i$ is the complex wave number of a resonance
($k_0>k_i>0$, and the symbol (*) means the complex conjugate
operation). Using Eq. (\ref{S-mat}), one can rewrite Eq.
(\ref{S-l}) in the form
\begin{equation}\label{S-phas}
S_l(k)=e^{2i(\nu_l+\delta_r+\delta_a)},
\end{equation}
where $\delta_r=-\arctan{\frac{k_i}{k-k_0}}$ stands for the
resonance phase shift, while
$\delta_a=-\arctan{\frac{k_i}{k+k_0}}$ is an additional phase
shift which  contributes to the whole scattering phase shift. Thus
the total phase shift is
\begin{equation}\label{phase}
\delta_l=\nu_l+\delta_r+\delta_a.
\end{equation}
The partial scattering nonresonant phase shift $\nu_l(k)$ is a
smooth function near the pole of the $S$-matrix element,
corresponding to the resonance. The $S$-matrix element defined by
Eq. (\ref{S-mat}) fulfills the conditions of analyticity, unitarity,
and symmetry. Therefore we can expand $\nu_l(k)$ to a series
\begin{equation}\label{series}
\nu_l(k)=\sum\limits_{n=0}^{\infty}c_n(k-k_s)^n
\end{equation}
in the vicinity of the pole corresponding to the resonance. The
point $k_s$ denotes a centered point, and a convergence radius
should be shorter than the distance from the centered point to the
closest singular point. The expansion coefficients $c_n$ of Eq.
(\ref{series}) as well as $k_0$ and $k_i$ (i.e., a resonance
complex energy value) are determined by  fitting the experimental
values of the elastic scattering phase shifts $\delta_l$ given by
Eq. (\ref{phase}). It is enough for the $\alpha\alpha$ system to
limit the expansion (\ref{series}) up to $n=4$. The nonresonant
phase shift $\nu_l(k)$ is an analytical function excluding the
origin. In Ref.  \cite{mur83}, the authors present the behavior
of $\delta_l(k)$ near the origin as
\begin{equation}\label{nu-l}
\delta_l(k)=-\frac{2\pi}{(l!)^2}k^{2l+1}\eta^{2l+1}a_le^{-2\pi\eta},
\end{equation}
where $a_l$ is the scattering length for colliding nuclei.
\footnote{There is a misprint in our paper \cite{irgaz}: In the
formula above $\nu$ was written instead of $\delta$ \cite{mur83}
but it does not matter because $\delta_r$ and
$\delta_a$ do not have such singularity.} The point  $k=0$ is  an
essential singularity point of the scattering phase shift.
However, as a function of the momentum $k, \nu$  has normal
analytical properties near the point corresponding to the
resonance. Besides,  the convergence region of Eq. (\ref{series}) is  limited due to the presence of an exchange Feynman diagram for the elastic
scattering, leading to the logarithmic singularity which is absent
in our models. The renormalized   partial amplitude is
constructed  (see Refs \cite{hamilton,orlov,akram84}) for its analytical continuation to a resonance or
bound-state energy region. According to its definition,  the
nuclear renormalized vertex constant $\tilde{G}_l$ (NVC) \cite{blok77},  NVC
can be written as
\begin{eqnarray}\label{NVC-2}
\tilde G_l^2&=&\frac{2\pi }{\mu^2}\frac{k_rk_ie^{i2\nu_l(k_r)}}{k_0\rho_l(k_r)}\nonumber\\
&=&\frac{\pi\Gamma}{\mu k_0}\frac{(1-ik_i/k_0)e^{i2\nu_l(k_r)}}{\rho_l(k_r)}.
\end{eqnarray}
where $\rho_l$ is equal to
\begin{equation}\label{rho}
\rho_l(k)=\frac{2\pi\eta}{e^{2\pi\eta}-1}\prod_{n=1}^l\Bigl(1+\frac{\eta^2}{n^2}\Bigr).
\end{equation}
Using the relationship between NVC $\tilde{G}_l$ and ANC $C_l$
\cite{blok77}, we obtain
\begin{eqnarray}\label{ANC}
C_l&=&\frac{i^{-l}\mu}{\sqrt{\pi}}\frac{\Gamma(l+1+i\eta_r)}{l!}e^{-\frac{\pi\eta_r}{2}}\tilde G_l\qquad\nonumber\\
&=&i^{-l}\sqrt{\frac{\mu\Gamma}{k_0}}e^{-\frac{\pi\eta_r}{2}}\frac{\Gamma(l+1+i\eta_r)}{l!}\nonumber\\
&\times&e^{i\nu_l(k_r)}\sqrt{(1-ik_i/k_0)/\rho_l(k_r)}.\qquad
\end{eqnarray}

In the limit of a small width $\Gamma$, we obtain the formula
\begin{equation}\label{narrow}
C^a_l=\sqrt{\frac{\mu\Gamma}{k_0}}e^{i(\nu_l(k_0)+\sigma_l(k_0)-\pi l/2)},
\end{equation}
which coincides with the expression derived in Ref.
\cite{akram77}. This formula (\ref{narrow}) can be used to check the calculation results. All the necessary expressions for the EFE are published  in the literature (see Ref. \cite{irgaz} and
references therein).

\section{The ground $^8\rm{Be}$ $\textbf{\it{s}}$ wave state resonance in the $\alpha\,\alpha$ scattering} \label{alpha-alpha}

In the present paper we continue to study resonances for light
nuclei. We consider the nucleus $^8\rm{Be}$ which is not bound in
the ground state due to the Coulomb repulsion between $\alpha$-particles. This state presents a very narrow resonance with the
pole at the center-of-mass system (c.m.s.) energy (see the review
\cite{tilley} and the references therein):
$$E_\alpha=E_0 - i\Gamma_0/2,\, E_0=91.84\, \rm{keV},\,
\Gamma_0= 5.57\pm 0.25\, \rm{eV}.$$
The $\alpha$-particle model is a good approximation for a
description of  $^{8}\rm{Be}$ characteristics because of the
$^{4}$He nucleus large binding energy. The other channels have
thresholds situated at $E_{lab}> 35$ MeV. The $Q$ value for the
reaction $\alpha+\alpha\rightarrow \,^{8}\rm{Be}$ has changed over time in the literature.
The value $Q = 94.5\pm 1.5$ keV is found in Ref. \cite{jones}.
Fowler's experimental group obtains $Q = 93.7 \pm 0.9$ keV
\cite{cook}. In Ref. \cite{russel} the results of the phase shift
analysis (see references in Ref. \cite{russel}) are used to find
the $^{8}\rm{Be}$ resonance parameters by applying the EFE. In
Ref. \cite{russel} the values of the energy $E_0$ and the width
$\Gamma_0$ for narrow resonance of $^8\rm{Be}$ in the ground state
in the cms frame are as follow:
$$
 E_0=94.5 \pm 1.4\,\rm{keV},\,\,\,
\Gamma_0= 4.5\pm 3\,\, \rm{eV}.
$$
The $E_0$ value is determined in Ref. \cite{russel} as the energy when
the phase shift $\delta_0$ passes $\pi/2$ and the resonance
width $\Gamma_0$ is obtained from the equation expressed $\Gamma$
in terms of the rate of changing $\delta_0$ in the resonance
region ($\pi/4 <\delta_0 <3\pi/4$). The nuclear
interaction is revealed in the scattering cross section at energy
$E_0> 300$ keV, i.e., after the resonance region where the
$s$-wave phase shift $\delta_0$ jumps from 0 up almost to
$\pi$. For an analytical continuation of the cross section into
the resonance region, the authors of  Ref. \cite{russel} apply the EFE
with the Coulomb interaction taken into account, using the formula
by Landau and Smorodinsky (see Ref. \cite{landau} which is valid in the
physical energy region and the reference to the original paper).
In  Ref. \cite{russel} the effective-range function $K_l(k^2)$ is
expanded in a series over $k^2$ up to power of $k^4$:
\begin{equation}\label{Kefrad}
K_l(k^2) = -1/a + (r/2)k^2 - P r^3 k^4,
\end{equation}

or an equivalent expansion in a series over $E_\alpha$
\begin{equation}\label{KefradE}
K_l(E_\alpha) = a_0 + a_1 E_\alpha + a_2 E_\alpha^2,
\end{equation}

which adequately describes the experimental values of $\delta_0$ at the
cms energy $E_\alpha\cong 2.5$ MeV. Later in Ref. \cite{benn} an uncertainty
of the measured cross-section from the Coulomb (Mott)
cross-section is found experimentally even in the resonance
region. As in Ref. \cite{russel}, the scattering of the singly-charged
ion $^4\rm{He}^+$ on the neutral $^4\rm{He}$ atoms is
investigated. The following results for $E_0$ and $\Gamma_0$ are
obtained in Ref. \cite{benn}:
\begin{equation}\label{Erbenn}
E_0 = 92.12 \pm 0.05\,\, \rm{keV}, \,\,\,\,\Gamma_0 = 6.8 \pm 1.7\,\,\rm{eV}.
\end{equation}
In later experiments (see \cite{russel}), the $E_0$ and $\Gamma_0$
values do not change much compared with  (\ref{Erbenn}),  but
their uncertainties appreciably decrease. The scattering
amplitude is defined in \cite{benn} as the sum of the Coulomb and the nuclear
amplitudes. The nuclear amplitude is taken in the Breit-Wigner
form, which may be a reasonable approximation for a narrow resonance.

In our paper we use the EFE, Pad\'e approximant, and SMP. The last
method is used for defining a resonant energy,
including that of broad resonances \cite{akram09}. In the case of a
narrow resonance, its energy depends less on the used method.
We show below that our result for the $^{8}\rm{Be}$  ground state
energy is in good agreement with (Eq. \ref{Erbenn}). As input
data, we use the phase shift borrowed from Ref. \cite{afzal} (see table
II in Ref. \cite{afzal}, p. 252). We use the following values for the
resonance energy and width (see Ref. \cite{tunl})
\begin{equation}\label{tunl}
E_0=91.84\,\, \rm{keV}; \,\,\,\Gamma_0 = 5.57\,\,
\rm{eV}.
\end{equation}
\begin{figure*}[thb]
\begin{center}
\parbox{12.0cm}{\includegraphics[width=12.0cm]{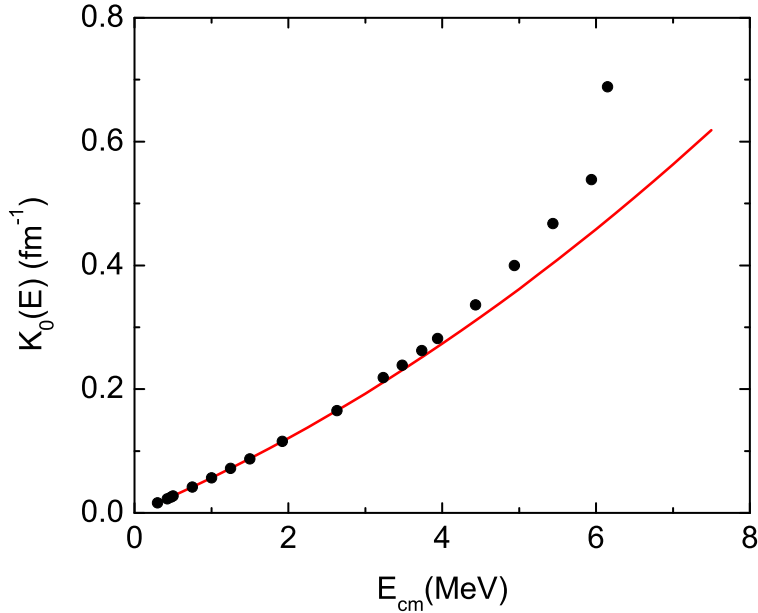}}
\end{center}
\caption{Comparison of the fitted effective-range function for
the $s$ wave $\alpha\alpha$ elastic scattering  with the experimental values. The
experimental data are taken from Ref. \cite{afzal}. The energy
is given in the c.m.s. frame.  \label{fig1}}
\end{figure*}
In Fig. \ref{fig1} we show the results of fitting the effective-range
function dependence on energy for the $s$ state of $^8\rm{Be}$ using
the phase shift table given in Ref. \cite{afzal}. It was indicated
earlier in Ref. \cite{OrlNik} that the radius of convergence of
the expansions in Eq. (\ref{Kefrad}) around zero
energy is determined by the singularity of the two-pion-exchange
Feynman diagram at an energy of $E_{sing} \approx 5$ MeV (more
precisely, at 4.76 MeV), since one-pion exchange is suppressed in
isospin. The fitting curve corresponds to the effective-range
function approximation (\ref{Kefrad}) with the parameters:
 \begin{equation}\label{K0 fitting}
a=-1724.1 \,\,\rm{fm};\,\,\, {\it r} =1.0848\,\, \rm{fm};\,\,\, {\it{P}}=-0.34717.
\end{equation}
This set does not differ much from that given in Ref. \cite{russel}:
\begin{equation}\label{K0 russel}
a=-1760\,\, \rm{fm}; \,\,\,{\it r} =1.096 \,\,\rm{fm};\,\,\, {\it{P}}=-0.314.
\end{equation}
We note that in Ref. \cite{russel}, misprints are made in writing
the shape parameter $P$: A dimension (cm) of $P$ was shown which
actually is a dimensionless quantity, and the wrong sign $P$ =
+0.314 is written (see the top part of Fig. 9 in Ref. \cite{russel}).
These misprints are not repeated in the bottom part of Fig. 9 where
the correct formula is written for the energy dependence of the
effective-range function $K_0(E_{lab})$. Furthermore, a reasonable
description of the experimental dependence $K_0(E_{lab})$ is shown
in the same figure in the area $E_{lab} \leq 6$
MeV (i.e., $E_\alpha \leq  3$ MeV in c.m.s. frame) with the parameters
(\ref{K0 russel}). For the $^8\rm{Be}$ ground state with the set (\ref{K0 fitting}) we receive the following values of the resonant energy,
width, NVC and ANC:
\begin{eqnarray}\label{EFE results}
E_0 &=&  92.248\,\, \rm{keV},\,\,\,\,\, \Gamma_0 = 5.122 \,\,\rm{eV};\,\,\,\,\,\nonumber\\
\tilde G_0^2&=&(0.5047 -i 0.0001151)\,\,  \rm{fm};\,\,\,\,\,\\
\mid \rm{ANC}\mid &=&\mid C_0\mid =0.001615\,\, \rm{fm}^{-1/2}.\nonumber
\end{eqnarray}
All the results in Eqs. (\ref{EFE results}) have complex values due
to the pole energy complexity. The resonance energy and width in
Eqs. (\ref{EFE results}) agree well with Eq. (\ref{tunl}). The
value of $\tilde G_0^2$ is almost real because of the resonance
pole proximity to the real energy axis. Using the expression
(\ref{narrow}) for narrow resonances we obtain $\mid C_0\mid = 0.001615\,\,\rm{fm}^{-1/2}$, which actually coincides with the value in Eqs. (\ref{EFE results}). This fact is quite natural for a
resonance with such a small $\Gamma_0$. When  $\Gamma_0$ is very small
and energy $E_0$ is situated not far from a threshold, one does not
need to use another method (for example, the SMP) because the related
results are practically the same as those for the explicit
expression (\ref{narrow}).

\section{The excited $^8\rm{Be}$: The $d$ wave state resonance in the $\alpha\,\alpha$ scattering} \label{alpha-alpha d state}

Along with a description of the $^{8}\rm{Be}$ ground state, the
survey in Ref. \cite{tilley} gives some features of the first excited
state of the $^{8}\rm{Be}$ nucleus ($J=2^+$). The amplitude for
the $\alpha\alpha$ scattering in the $d$ wave has a pole at a
cms energy of $E_\alpha = E_{\rm{cm}} = E_2 - i\Gamma_2/2$ (see Table (8.11) in Ref.
\cite{tilley} and notes to it on p. 184). In the laboratory
frame, the respective energy is $E_{lab} = 2E_{\rm{cm}} (E_{\rm{cm}} = k^2/2\mu$,
where $k$ is the relative momentum of colliding $\alpha$ particles
and $\mu$ is the reduced mass of the $\alpha\alpha$ system). The
weighted-mean values of the real part of the energy at the pole,
$E_2 \pm \Delta E_2$, and of the resonance width, $\Gamma_2 \pm
\Delta\Gamma_2$, are given in Ref. \cite{tilley} along with the
respective inaccuracies:
\begin{equation}\label{Dresonance}
E_2 = 3.03 \pm 0.01\,\, \rm{MeV},\,\,\, \Gamma_2 = 1.49 \pm 0.02 \,\,\rm{MeV}.
\end{equation}
These values are found from later results on the yields of the
reactions $^9{\rm{Be}}(p, d)$ and $^9{\rm{Be}}(d, t)$ (see also Ref.
\cite{ajzenberg}). Similar results of analyses performed for
various reactions by various groups of authors at various times
are also given in Ref. \cite{tilley} [see the references in Table (8.9)]. Those results differ only in the value of the uncertainty in the resonance width, $\Delta\Gamma_2 = 0.015$ MeV. From Table
(8.11) in Ref. \cite{tilley}, we find the ranges of mean values of
$E_2$ and $\Gamma_2$ and the scatter of the uncertainties in them:
\begin{eqnarray}\label{Dresonance_scatter)}
2.82&\leq& E_2 \leq 3.18 \,\,\rm{MeV},\,\,\, 10 \leq \Delta E_2 \leq 200\,\,\rm{keV};\nonumber\\
1.20 &\leq&  \Gamma_2 \leq 1.75 \,\,\rm{MeV},\,\,\,
20 \leq \Delta\Gamma_2 \leq 300\,\,\rm{keV}.
\end{eqnarray}
One can see that $E_2$ and $\Gamma_2$ and the boundaries of their
variations are commensurate. It is noteworthy, however, that
$\alpha\alpha$ scattering is not present among the reactions
appearing in Table (8.11) from Ref. \cite{tilley}. One of the
objectives of the present study is to supplement the data
quoted in Ref. \cite{tilley} with data on $\alpha\alpha$ scattering by
using the effective-range theory and $S$-matrix pole method. For
the $d$ wave resonance in question, we present the
effective-range function in the form (\ref{Kefrad}) of an
expansion in the powers of $k^2$ up to $k^4$, and in
the form  which is the equivalent of the expansion in (\ref{KefradE}):
\begin{equation}\label{KEefrad}
K_2(E_\alpha) = A_2 +B_{21} E_\alpha + B_{22} {E_\alpha}^2.
\end{equation}
The calculated values of the functions $K_2(E_\alpha)$ and
$\delta_2(E_\alpha)$ are highly sensitive to the position of
the pole in the complex energy plane, and especially to the value
$E_2$. The use of the values in Eq. (\ref{Dresonance}) in fitting
the parameters of the effective-range function $K_2(E_\alpha)$
along with the experimental value of the phase shift $\delta_2$
from Ref. \cite{afzal} (see Table II there) at the energy where the
uncertainty $\Delta\delta_2$ is minimal distorts the shape of the
energy dependence in relation to the experimental data. A partial-wave
phase shift analysis of $\alpha\alpha$ scattering was performed
more than 40 years ago (see references in Ref. \cite{afzal}). The more
recent publication by Warburton \cite{warburton} contains
information on the experimental dependence $\delta_2 (E_α)$ in
the form of a graph. There is virtually no difference between the
data in Ref. \cite{warburton} and the data on $\delta_2 (E_\alpha)$ in Ref.
\cite{afzal} (Table 2 on p. 252), with the exception of
several extra points in the region where $\delta_2\leq \pi/2$. The
positions of these points fit well with the general character of the
energy dependence of $\delta_2$ in Ref. \cite{afzal}. The phase shift
$\delta_2$ begins to manifest itself for $E_\alpha \geq$ 1.25 MeV, the
resonance lying completely in the region of convergence of
$K_2(E_\alpha)$.  Indeed, we find from Eq. (\ref{Dresonance}) that $\mid E_2
- i\Gamma_2/2\mid \approx$ 3 MeV $<$ 5 MeV. As soon as the
considered resonance is broad enough, we apply both methods (the EFE
and the SMP) to describe its characteristics. We take the experimental
\cite{tunl} resonance energy and width as
\begin{equation}\label{tunl2}
E_2 =  3.122 \,\,\rm{MeV},\,\,\, \Gamma_2 = \,\,1.513 \rm{MeV}.
\end{equation}
First, we use the EFE method. The corresponding fitting
effective-range function $K(E_\alpha)$ taken in the form
(\ref{KEefrad}) leads to $K(E_\alpha)$
parameters (with an energy in MeV):
\begin{eqnarray}\label{KEFEParE}
A_2 &=& 0.0182 \,\,\rm{fm}^{-5},\,\,\, {\it B}_{21} = -0.0056\,\, \rm{fm}^{-5}\rm{MeV}^{-1},\,\,\,\,\nonumber\\
B_{22}&=& 0.0027\,\, \rm{fm}^{-5}\rm{MeV}^{-2}.
\end{eqnarray}
The set (\ref{KEFEParE}) corresponds to the parameters of Eq. (\ref{Kefrad}):
\begin{equation}\label{KEFEPark}
a = - 55.0\,\, \rm{fm}^5, \,\,\,{\it r} = -0.1166\,\, \rm{fm}^{-3},\,\,\, {\it P} = 183.9\,\, \rm{fm}^8.
\end{equation}
\begin{figure*}[thb]
\begin{center}
\parbox{12.0cm}{\includegraphics[width=12.0cm]{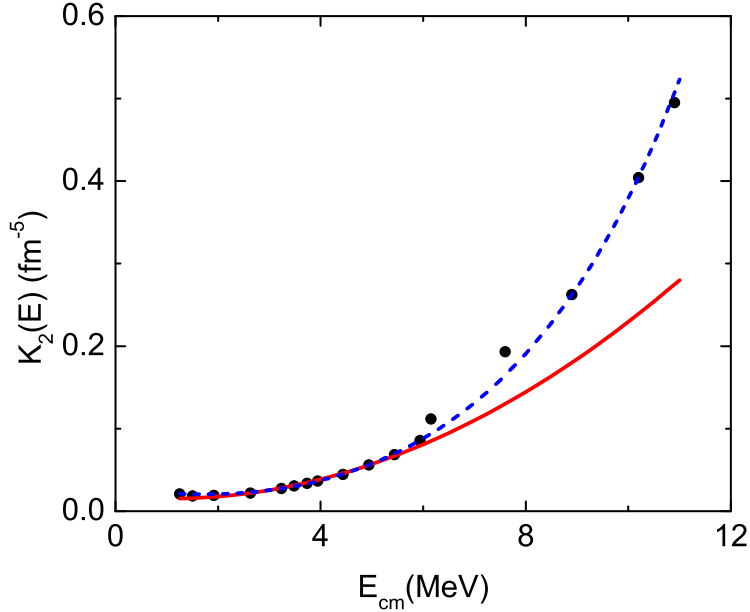}}
\end{center}
\caption{Comparison of the fitted effective-range function for
the $d$ wave $\alpha\alpha$ elastic scattering (red color line) with the experimental values (points). The dashed line represents the result of fitting with the Pad\'e approximant. The
experimental data are taken from Ref. \cite{afzal}. The energy
is given in the c.m.s. frame.  \label{fig2}}
\end{figure*}
The function $K_2(E_\alpha)$ fitting results are shown in Fig. \ref{fig2}
(a solid curve) where the  experimental  energy interval up to 11
MeV (in the c.m.s. frame) is considered. One can see that this EFE variant
describes the experimental phase shift dependence on the energy
only when $E_{lab}< 5$ MeV. We get the following results:
\begin{eqnarray}\label{EFE2results}
E_2 &=& 2.897 \,\,\rm{MeV},\,\,\, \Gamma_2 = 1.470\,\, \rm{MeV};\nonumber\\
\tilde G_2^2 &=& (0.0137-i 0.0169) \,\,\rm{fm};\\
\mid \rm{ANC}\mid &=& \mid C_2\mid = 0.3152 \,\,\rm{fm}^{-1/2}.\nonumber
\end{eqnarray}
The resonant energy $E_2$ and the $\Gamma_2$ in Eqs. (\ref{EFE2results}) are
smaller than the experimental values given in Ref. \cite{tunl} but
the differences are not large. We also note that when
using Eq. (\ref{narrow}) (for a narrow resonance) $|C_2|$ = 0.3624 fm$^{-1/2}$. The
difference is not very big for such a broad resonance.
To extend a good description of $K_2(E_\alpha)$  when $E_\alpha >5$
MeV, we apply a Pad\'e approximant, adding one more
parameter to the effective-range function, which takes the form:
\begin{equation}\label{KEPade}
K_2(E_\alpha)=[a_0 + a_1 E_\alpha + a_2{E_\alpha}^2]/(1+b_1E_\alpha),
\end{equation}
where
\begin{eqnarray}
a_0&=&0.028\,\, \rm{fm}^{-5},\,\,\, {\it a}_1=-0.009799 \,\, \rm{fm}^{-5}\rm{MeV}^{-1},\\
a_2&=&0.002549 \,\,\rm{fm}^{-5}\rm{MeV}^{-2},\,\,\, {\it b}_1=-0.05122\,\,\rm{MeV}^{-1}.\nonumber
\end{eqnarray}

In Fig. \ref{fig2} the dashed curve for Eq. (\ref{KEPade}) practically
coincides with the solid curve at $E_\alpha\leq 5 $ Mev but also reproduces
 well the experimental  points at $E_\alpha\geq 5$ Mev. We obtain the
following results:
\begin{eqnarray}\label{Pade2results}
E_2 &=& 2.9380\,\, \rm{MeV}\,\,\, \Gamma_2 = 1.2296 \,\,\rm{MeV};\nonumber\\
\tilde G_2^2 &=& (0.0117 -i 0.0136) \,\,\rm{fm};\\
\mid \rm{ANC}\mid&=&\mid C_2\mid =0.289\,\, \rm{fm}^{-1/2}.\nonumber
\end{eqnarray}
The differences between the related values in Eqs. (\ref{EFE2results}) and
(\ref{Pade2results}) are not very large. One can see
that the resonance energy and width are smaller than the experimental
values (\ref{tunl2}).

\begin{figure*}[thb]
\begin{center}
\parbox{12.0cm}{\includegraphics[width=12.0cm]{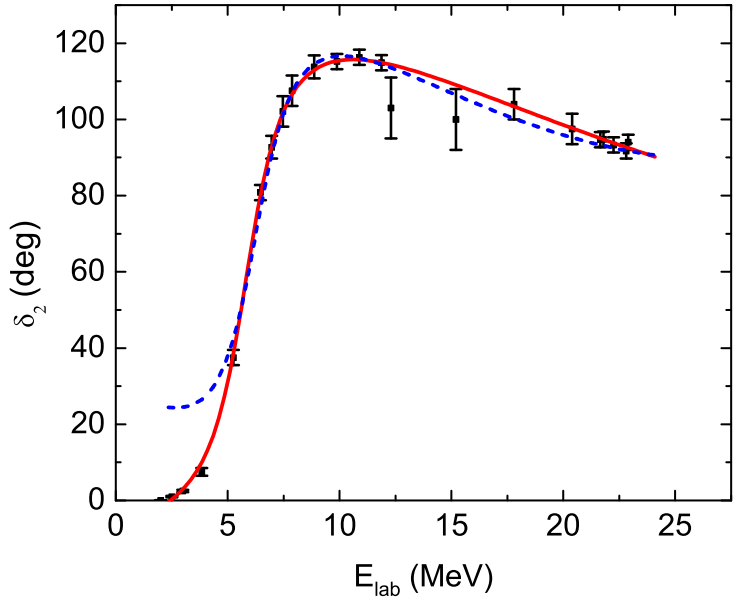}}
\end{center}
\caption{Comparison of the fitted phase shifts for
the $d$ wave $\alpha\alpha$ elastic scattering obtained by the
$S$-matrix pole method  with the experimental data taken from Ref. \cite{afzal}. The solid line shows when the complex momentum $k_r$ of the resonance is added to the parameters of fitting. The dashed line indicates when the energy and width of resonance is fixed as in Eq. (\ref{tunl2}). The energy
is given in the laboratory frame.  \label{fig3}}
\end{figure*}
We apply the $S$-matrix pole method to the excited $2^+$
state. We use the experimental resonance energy (\ref{tunl2}) as a trial value to
estimate the centered momentum value for the series expansion
(\ref{series}). A very good phase shift  fitting  is achieved with
four members up to $(k-k_s)^3$  in Eq. (\ref{series}). The model
describes the phase shift behavior as a function of the energy in the
whole energy region (see Fig. \ref{fig3}) with the exception of the two
experimental points with large uncertainties which obviously
disagree with the general  trend of the phase shift energy
dependence. We find the following results:
 \begin{eqnarray}\label{SMPresults}
E_2 &=& 2.916\,\, \rm{MeV},\,\,\,\Gamma_2 = 1.437\,\,\rm{MeV};\nonumber\\
\tilde G_2^2 &=& (0.01537 - i0.0145)\,\,\rm{fm};\\
\mid \rm{ANC}\mid &=&\mid C_2\mid = 0.3120\,\, \rm{fm}^{-1/2}.
\end{eqnarray}
We also perform a fitting  parameters of the $S$-matrix pole method with the fixed values of the energy and width of the $d$ state resonance, which are generally accepted as the experimental values and which are given in Eq. (\ref{tunl2}). In this case we also use the expansion up to terms $(k-k_s)^3$ of the nonresonant phase shift. However, as can be seen from Fig. \ref{fig3} (dashed line), the agreement with the curve of the experimental points is worse compared with the previous method of the fitting, when we take the real and imaginary momentum of the resonance as the fitting parameters. We find the following results for the NVC and ANC:
\begin{eqnarray}\label{SMP-fixresults}
\tilde G_2^2 &=& (0.0160 - i0.0067)\,\,\rm{fm};\nonumber\\
\mid \rm{ANC}\mid &=& \mid C_2\mid = 0.2914\,\, \rm{fm}^{-1/2}.
\end{eqnarray}

One can see from Eqs. (\ref{EFE2results}), (\ref{Pade2results}),  (\ref{SMPresults}) and (\ref{SMP-fixresults})  that the results are very
sensitive to the method used for fitting.  In spite of the good
phase shift fitting, we find that the different low-energy approaches lead to
quite different results. The resonant energy occurs to be especially
changeable and  varies. This means there is disagreement
between the experimental energy dependence of the phase shift on the one
hand  and the resonance complex energy  on the other. So we recommend
 refining the phase shift for $\alpha\alpha$ scattering.
Nevertheless some  estimations of the ANC found in this
work can be used in astrophysics.

\section{$^{16}\rm{O}$ bound states properties from the $\alpha^{12}\rm{C}$ scattering phase shifts} \label{alpha-12C}

The SMP is applicable to resonances but not to bound states. So in
this section we apply the effective-range theory and the
Pad\'e approximant. For all considered $^{16}\rm{O}$ bound states, the fitted
effective-range function is quite well reproduced when the expansion (EFE) is
limited to the expression (\ref{Kefrad}) or (\ref{KEefrad}) (with
the dependence on $k^2$ or $E_{\rm{cm}}$). For the ground state $J^\pi=0^+$,
the fitting  curve for $K_0(E_{\rm{cm}})$  is almost linear with the set
of parameters in Eq. (\ref{KefradE}):
\begin{eqnarray}\label{0coef}
a_0&=&-0.000328\,\rm{fm}^{-1};\,\, {\it a}_1=0.019450\,\rm{fm}^{-1}\rm{MeV}^{-1};\nonumber\\ a_2&=&0.000171936\,\rm{fm}^{-1}\rm{MeV}^{-2}.
\end{eqnarray}
The curve $K_0(E_{\rm{cm}})$ crosses the point corresponding to the ground
state at the experimental binding energy $\varepsilon=7.162$ MeV. The
fitting parameters (\ref{0coef}) lead to the following results:
\begin{equation}\label{0results16O}
{\tilde G_0}^2 = 5.197\,\, \rm{fm};\,\,\, \rm{ANC}=20.33\,\, \rm{fm}^{-1/2}.
\end{equation}
For the excited state $J^\pi=1^-$ with the binding energy $\varepsilon_1=0.045$ MeV, we find the following fitting set of parameters:
\begin{eqnarray}\label{1coef}
a_0&=&-0.00001773 \,\rm{fm}^{-3}; \,\,{\it a}_1=0.02930 \,\rm{fm}^{3}\rm{MeV}^{-1};\nonumber\\
a_2&=&0.003321\, \rm{fm}^{-3}\rm{MeV}^{-2}.
\end{eqnarray}
The curve $K_1(E_{\rm{cm}})$ crosses the point $E_{\rm{cm}}=-0.045$ Mev corresponding to the
excited $1^-$ state almost at the the point corresponding to the
experimental binding energy and describes the experimental points quite well. The set of parameters  (\ref{1coef}) leads to the following results:
\begin{equation}\label{1results16O}
{\tilde G_1}^2 = 0.03584\,\rm{fm};\,\, \rm{ANC}=1.032\times10^{14}\, \rm{fm}^{-1/2}.
\end{equation}
The very large ANC value for $1^-$ state is due to the small
binding energy for this subthreshold level. For the excited state
$J^\pi=2^+$ with the binding energy $\varepsilon_1=0.245$ MeV,
we find the following fitting set of parameters:
\begin{eqnarray}\label{2coef}
a_0&=&-0.0004657\,\,\rm{fm}^{-5};\,\, {\it a}_1=0.00993814\,\rm{fm}^{-5}\rm{MeV}^{-1};\nonumber\\
a_2&=&0.007050\,\, \rm{fm}^{-5}MeV^{-2}.
\end{eqnarray}
The curve $K_2(E_{cm})$ crosses the point corresponding to the
excited $2{^+}$ state almost at the point $E_{cm}=-0.245$ MeV, corresponding to the
experimental binding energy, and describes the experimental points reasonably well. The set of parameters (\ref{2coef})
leads to the following results:
\begin{equation}\label{2results16O}
{\tilde G_2}^2 = 0.001183\,\rm{fm};\,\, ANC=21060\,\rm{fm}^{-1/2}.
\end{equation}.

\section{$^{16}\rm{O}$ resonant-state properties from the $\alpha^{12}\rm{C}$ scattering phase shifts found using the Pad\'e approximant for the effective-range function} \label{alpha-12Cpade}

In our paper \cite{irgaz} we find that the
effective-range method (EFE) is not able to reproduce properly the
widths of the $^{16}\rm{O}$ resonances while the $S$-matrix pole
method can give reasonable results. In Ref. \cite{irgaz} we conclude
that the SMP is successful because the central point $k_s$ for expansion
(\ref{series}) is  situated just near the resonance pole. In this
section of the present paper we try to improve this situation by
applying the Pad\'e-approximant for the effective-range
function instead of the  polynomial expansion (EFE), taking into account
the fact that the Pad\'e approximant better reproduces the
phase shift energy dependence. To do this, we
study the resonances for the states with $J^\pi=1^-$ and
$3^-$ where the widths are bigger compared with other
states (see Table III in Ref. \cite{irgaz}). The SMP fits the experimental phase shifts for these states  quite well. Fig. 3 of Ref. \cite{irgaz} shows that the resulting curves actually cross the experimental points.
In Table \ref{tab1} we compare the respective results obtained by the SMP and
Pad\'e approximant. Again we see that the results  are not very
different.
\begin{table*}[!ht]
\caption{States, energies and widths of $^{16}\rm{O}$ nucleus
levels above the $\alpha^{12}\rm{C}$ threshold from our fits, as
well as the corresponding values of the calculated NVC and ANC
from the elastic $\alpha^{12}\rm{C}$ scattering phase shifts
\cite{tisch}. Four terms of Eq. (\ref{series}) are used for
fitting. The energies of the resonances are given in the
center-of-mass system of $\alpha^{12}\rm{C}$}.
\begin{ruledtabular}

\begin{tabular}{lcccrcc}
&&&&&&\\
Method&$J^\pi$&$E_r$\,(MeV)&$\Gamma$\,(MeV)
&$\tilde{G}^2_l$\,(fm)\quad\,\,&$C_l\,(\rm{fm}^{-1/2})$&$\mid C_l\mid\,(\rm{fm}^{-1/2}$)\\
&&&&&&\\
\hline
&&&&&&\\
SMP&$1^-$&2.364& 0.356 &$4.970-i1.797$&$0.1530-i0.1032$&$0.185$\\
Pad\'e&$1^-$&2.362& 0.347&$4.862-i1.789$&$0.1516-i0.1015$&$0.182$\\
SMP&$3^-$&4.214&0.812&$0.276-i0.142$&$-0.233-i0.020$&$0.234$\\
Pad\'e&$3^-$&4.239&0.786&$0.270-i0.112$&$-0.228-i0.029$&$0.230$\\
 \end{tabular}
 \end{ruledtabular}
\label{tab1}
\end{table*}

\section{Conclusion}\label{conclusion}

In the present paper we continue to study the resonant states
of light nuclei. Concretely we consider  $^8\rm{Be}$ in the $0^+$ and
$2^+$ states and $^{16}\rm{O}$ in the $1^-$ and $3^-$ states using
three different low-energy methods: the effective-range expansion, the
Pad\'e approximant and the $S$-matrix pole method to calculate ANCs.
We use as an input the phase shift energy behavior for
the $\alpha\alpha$ and $\alpha^{12}\rm{C}$ scattering borrowed
from the literature. The SMP method is not applicable to bound
states, so the EFE is used in the present paper  to obtain the ANC
for the three $^{16}\rm{O}$ bound states: one ground
($\varepsilon=7.162 \,\rm{MeV},\,\,J^\pi=0^+$) and two excited
($\varepsilon=0.245,\,\,\rm{MeV},\,\,J^\pi=2^+$) and ($\varepsilon=0.045\,\,\rm{MeV},\,\,J^\pi=1^-$).
All the nuclei considered are very important in astrophysics. But until
now their ANC (at least for resonance states) have not been estimated
theoretically. We emphasize that our
methods for finding the resonant states  ANC allow us to normalize the
Gamov wave function which is quite difficult,
especially in the presence of the Coulomb interaction. All the  methods
considered here are based on the phase shift analysis and on the
analytical continuation of elastic scattering amplitudes
(renormalized due to the Coulomb interaction) to nonphysical
energy region. The results of this
work show the contradiction between the resonance energy and the
phase shift energy dependence. As the phase shift
data for $\alpha\alpha$ scattering is about 40 years old, we
recommend remeasuring. From our calculations we draw the following conclusions.

The $^8\rm{Be}$ ground $0^+$ resonant state is so narrow that the
results are not sensitive to the model applied. In this case, a simple
expression (\ref{narrow}) can be used for the ANC, which is developed in Ref. \cite{akram77}.

The resonance for the $^8\rm{Be}$ excited $2^+$ state is broad enough to
be more sensitive to the model used. We find in Ref. \cite{irgaz} that the EFE does not reproduce the $\Gamma$ value while the SMP leads to a reasonable result for $\Gamma$. We also find that the
application of the Pad\'e-approximant instead of the EFE shows a better
agreement with the SMP results. In this case, adding another
parameter to introduce the pole into the effective-range function (zero
for a partial scattering amplitude) leads to a much better
description of $K_2(E_\alpha)$ in a larger energy area.

We show that applying the Pad\'e-approximant to the $^{16}\rm{O}$ in $1^-$ and $3^-$
states resonances also improves the agreement with the SMP
results published in Ref. \cite{irgaz}. So in a way
effective-range methods are 'rehabilitated' in the present
paper after increasing the number of the fitting parameters and introducing a pole.
In spite of this, the SMP is a better method  for describing properties for broad resonances.
A very good description of the experimental phase shift data is shown in Fig. 3 of our previous paper \cite{irgaz}.

The $^{16}\rm{O}$ bound states are also studied. We find
very large differences in the ANC depending on the binding energy. The
resonance near the threshold has a very big ANC due to the
$\Gamma$-function in its definition (\ref{ANC}). We note that in
Ref. \cite{spren} the authors calculate the ANC for the first
$2^+$ excited state of $^{16}\rm{O}$. They  choose a nuclear
Gaussian potential, which reads
$V(r)=-112.3319 \exp(-r^2/2.82)\, \rm{MeV}$,
where $r$ is the distance between the clusters in fm, and the screened Coulomb potential
is $e^2 {\rm{erf}}(r/2.5)/r$, where erf$(x)$ is the error function. This potential
has a bound state when $E_\alpha = -245.0$ keV.   Numerically, the authors
find the ANC  = $1.384 \times 10^5\,\rm{fm}^{-1/2}$. This value can be
compared with our result ANC=$1.0323\times 10^4\,\rm{fm}^{-1/2}$.

The results of this paper can be used for solving  nuclear astrophysical
problems and may be applied to the theory of nuclear reactions
using Feynman diagrams to describe the reaction mechanisms.

This work is supported by the Russian Foundation for Basic
Research (project No. 13-02-00399). We are grateful to
H.\,M.~Jones for editing the English of this manuscript.


\begin{thebibliography}{00}
\bibitem{imbriani} G.~Imbriani, M.~Limongi, L.~Gialanella, \textit{et al.}, Astrophys. J. \textbf{558}, 903 (2001).
\bibitem{rausher} T.~Rausher, A.~Heger, R.\,D.~Hoffman,S.\,E.~Woosley, Astrophys. J. \textbf{576},323 (2002).
\bibitem{straniero} O.~Staniero, I.~Dominguez, G.~Imbriani, L.~Piersanti, Astrophys. J. \textbf{583},878 (2003).
\bibitem{hill} W.~Hillerbrandt, J.~Niemeyer, Annu. Rev. Astron. Astrophys. \textbf{38},191 (2000).
\bibitem{irgaz} B.\,F.~Irgaziev, and Yu.\,V.~Orlov, Phys. Rev. C \textbf{91}, 024002 (2015)\textbf{.}
\bibitem{akram09} A.\,M.~Mukhamedzhanov, B.\,F.~Irgaziev, V.\,Z.~Goldberg, Yu.\,V.~Orlov, and I.~Qazi, Phys. Rev. C \textbf{81}, 054314 (2010).
\bibitem{Hoyle} F.~Hoyle, Astrophys. J. Suppl., \textbf{1}, 121 (1954).
\bibitem{migdal} A.\,B.~Migdal, A.\,M.~Perelomov, and V.\,S.~Popov, Yad. Fiz. \textbf{14},
874 (1971) [Sov. J. Nucl. Phys. \textbf{14}, 488 (1972)].
\bibitem{mur83} V.\,D.~Mur, A.\,E.~Kudryavtsev, V.\,S. ~Popov, Yad. Fiz., \textbf{37} 1417 (1983) [Sov. J. Nucl. Phys., \textbf{37}, 844 (1983)].
\bibitem{hamilton} J.~Hamilton, I.~{\O}verb\"o, B.~Tromborg, Nucl. Phys. \textbf{B 60}, 443 (1973).
\bibitem{orlov} Yu.\,V.~Orlov, B.\,F.~Irgaziev, L.\,I.~Nikitina, Yad. Fiz. \textbf{73}, 787 (2010) [Phys. At. Nucl. \textbf{73}, 757 (2010)].
\bibitem{akram84} L.\,D.~ Blokhintsev, A.\,M.~Mukhamedzhanov,  A.\,N.~Safronov, Fiz. Elem. Chastits At. Yadra, \textbf{15}, 1296 (1984) [Sov. J. Part. Nucl., \textbf{15}, 580 (1984)].
\bibitem{blok77} L.\,D.~Blokhintsev, I.~Borbely and E.\,I.~Dolinsky, Fiz. Elem. Chastits At. Yadra, \textbf{8}, 1189 (1977) [Sov. J. Part. Nucl., \textbf{8}, 485 (1977)].
\bibitem{akram77} E.\,I.~Dolinsky, and A.\,M.~Mukhamedzhanov, Izv. Acad. Nauk SSSR, Ser Fiz, \textbf{41}, 2055 (1977). [Bull. Acad. Sci. USSR, Phys. Ser. \textbf{41}, 55 (1977)].
\bibitem {tilley} D.\,R.~Tilley,  J.\,H.~Kelley, J.\,L.~Godwin,  \textit{et al.}, Nucl. Phys. {\textbf A, 745}, 155 (2004).
\bibitem{jones} K.\,W.~Jones, D.\,J.~Donahue, \textit{et al.}, Phys. Rev. 91, 879 (1953).
\bibitem{cook} C.\,W.~Cook, W.\,A.~Fowler, C.\,C.~Lauritsen, \textit{et al.}, Phys. Rev., \textbf{107}, 508 (1957).
\bibitem{russel}J.\,L.~Russell, Jr., G.\,C.~Phillips, and C.\,W.~Reich, Phys. Rev. 104, 135 (1956).
\bibitem{landau}  L.\,D.~Landau, E.\,M.~Lifshitz, \textit{Quantum Mechanics:
Nonrelativistic Theory}, 3rd ed. (Pergamon, Oxford, UK, 1977).
\bibitem{benn} J.~Benn, E.\,B.~Dally, and H.\,H.~Muller textit{et al.}, Nucl. Phys. \textbf{A, 106}, 296 (1968).
\bibitem{afzal} S.\,A.~Afzal, A.\,A.\,Z.~Ahmad,, and S.~Ali, Rev. Mod. Phys., \textbf{41}, 247 (1969).
\bibitem{tunl} http://www.tunl.duke.edu/nucldata/chain/08.shtml.
\bibitem{OrlNik} Yu.\,V.~Orlov,and L.\,I. ~Nikitina, Izv. Ross. Akad. Nauk, Se. Fiz. \textbf{76}, 503 (2012)  [Bull. Russ. Acad. Sci.: Phys., \textbf{76}, 446 (2012)].
\bibitem{ajzenberg} F.~Ajzenberg-Selove, Nucl. Phys. A, \textbf{320}, 1 (1979).
\bibitem{warburton} E.\,K. ~Warburton, Phys. Rev. C \textbf{33}, 303 (1986).
\bibitem{tisch} P.~Tischhauser, \textit{et al.}, Phys. Rev. C \textbf{79}, 055803 (2009).
\bibitem{spren} Jean-Marc ~Sparenberg, Pierre ~Capel, and Daniel ~Baye, Phys. Rev. C \textbf{81}, 011601(R) (2010).
\end{thebibliography}
\end{document}